\documentclass[preprint,superscriptaddress,amsmath,amssymb,prb,]{revtex4-2}
\usepackage{graphicx} % Required for inserting images
\usepackage{amsmath}
\usepackage{indentfirst}
\usepackage{multirow}
\usepackage{subfigure}
\usepackage{listings}
\usepackage{xcolor}
\begin{document}

\title{Dataset Distillation for Machine Learning Force Field in Phase Transition Regime}
\author{Ruiyang Chen}
 \thanks{These authors contribute equally to this work}
 \affiliation{School of Physics, Peking University, Beijing 100871, People's Republic of China.}
\author{Qingyuan Zhang}
 \thanks{These authors contribute equally to this work}
 \affiliation{School of Physics, Peking University, Beijing 100871, People's Republic of China.}
\author{Ji Chen}
 \email{ji.chen@pku.edu.cn}
 \affiliation{School of Physics, Peking University, Beijing 100871, People's Republic of China.}
 \affiliation{Interdisciplinary Institute of Light-Element Quantum Materials and Research Center for Light-Element Advanced Materials, Peking University, Beijing 100871, People's Republic of China}
 \affiliation{State Key Laboratory of Artificial Microstructure and Mesoscopic Physics and Frontiers Science Center for Nano-Optoelectronics, Peking University, Beijing 100871, People's Republic of China}

\date{\today}

\begin{abstract}

Machine learning force field (MLFF) has emerged as a powerful data-driven tool for atomistic simulations, enabling large-scale and complex atomic systems to be simulated with accuracy comparable to \textit{ab initio} methods. 
However, MLFFs often suffer from low training efficiency in the phase transition regime, where structural fluctuations are significantly elevated. To address this challenge, we propose a Central-Peripheral Distillation (CPD) algorithm for training dataset distillation.
By strategically integrating representative samples with critical corner cases, the CPD algorithm ensures that the distilled dataset retains maximum structural diversity. 
We validated the efficacy of the CPD method on the liquid-liquid phase transition of dense hydrogen. 
Results show that, with the CPD approach, only 200 configurations are sufficient to train a MLFF that can fully reproduce the structural and dynamical properties of liquid hydrogen in the vicinity of its phase transition regime.
This work paves the way for high-fidelity labeling of the MLFF training datasets, for instance by adopting high-level \textit{ab initio} calculations beyond the standard density functional theory, thereby enhancing the predictive accuracy of MLFFs.

%\medskip
%\noindent \textit{Keywords:} Machine Learning Force Fields, Data Distillation, CPD Algorithm, Phase Transition, Hydrogen
\end{abstract}

\maketitle
\addtocontents{toc}{\protect\setcounter{tocdepth}{-1}}

\section{Introduction}
%The accurate construction of molecular potential energy surfaces (PES) is a fundamental science in modern materials science. A PES not only defines the fundamental interaction patterns between atoms but also serves as an indispensable prerequisite for conducting molecular dynamics (MD) simulations. By performing dynamical simulations across the PES, researchers can capture the temporal evolution of a system, thereby bridging the gap between microscopic structural characteristics and macroscopic physical properties. Although ab-initio methods provide high-precision energy and force data, their computational cost severely constrains their application in large-scale, long-duration simulations of chemically complex systems.

In recent years, machine learning force field (MLFF) has emerged as a compelling solution for the accurate construction of potential energy surfaces (PESs). 
By fitting high-accuracy data derived from high-precision \textit{ab initio} calculations, MLFF models can reproduce the PES with \textit{ab initio} accuracy while maintaining superior computational efficiencies. 
This capability renders feasible large-scale, long-duration simulations of complex materials. 
A primary advantage of MLFFs lies in their ability to model highly complex configurations, providing deeper insights into diverse physicochemical processes \cite{bpnn_2007,Schtt2017SchNetA,wang2018deepmd,Batzner2021E3equivariantGN,Deringer2021GaussianPR,unke_mlff_2021,batatia2022mace}. 
However, the robustness and predictive accuracy of MLFFs are highly dependent on the quality and quantity of the training data. 
Efficiently identifying the most representative configurations from a vast, high-dimensional configurational space remains a critical bottleneck in the field \cite{unke_mlff_2021}. 
This difficulty is particularly pronounced when MLFFs are applied to study phase transition problems, where fluctuations enlarge the configuration space\cite{Jinnouchi2019OntheflyML,Jinnouchi2019PhaseTO,cheng2020evidence,Chen2025HighPM}.

\begin{figure*}[t]
    \centering
    \includegraphics[width=1\linewidth]{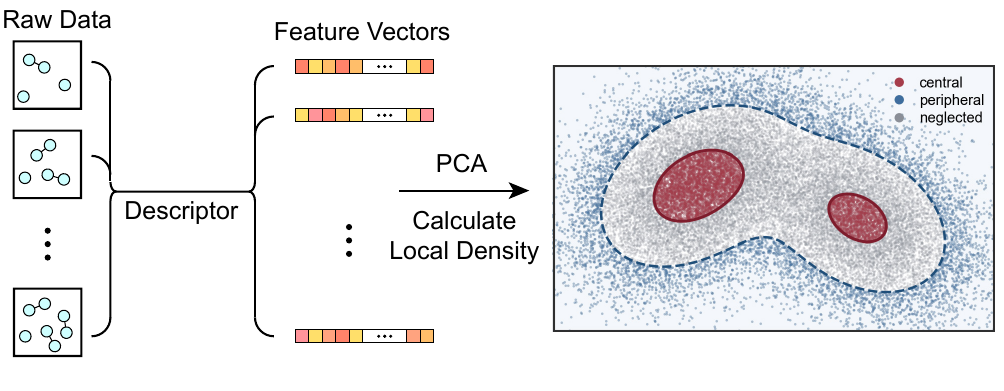}
    \caption{Schematic illustration of the CPD sampling workflow. The CPD algorithm extracts molecular features via MACE and PCA, followed by an optimized local density analysis. By employing a dual-focus weighted sampling strategy—targeting the top 20\% densest (central) and 20\% sparsest (peripheral) points, the model captures both representative phase characteristics and critical rare configurations, maximizing the structural diversity of the distilled dataset.}
    \label{fig1}
\end{figure*}
To improve the quality of training dataset for MLFF, two primary approaches have emerged in the field. 
The first category, represented by active learning, is centered on expanding the configurational diversity of training sets through structure generation algorithms, thereby enhancing the scalability of MLFF models \cite{Deringer2017DataDrivenLO,Smith2018LessIM,Sivaraman2020MachinelearnedIP,Vandermause2021ActiveLO}. 
The second approach, typified by configurational data distillation, focuses on eliminating redundant data from training datasets to improve the fitting efficiency of the models \cite{Finkbeiner2023GeneratingMT,Qi2023RobustTO,li2023exploiting,schwalbe2025model}.
Notably, the latter approach is particularly critical when further elevating the model’s accuracy to match higher-level \textit{ab initio} methods. 
This is because state-of-the-art \textit{ab initio} methods, such as coupled-cluster theories, quantum Monte Carlo, and neural network wavefunctions are computationally extremely expensive \cite{Booth2012TowardsAE,Hermann2022AbIQ,ren2023towards,Shi2024AnAA,tang2025deep,qian2025deep,huang2025multi}. 
Although, currently the primary data source for MLFF training is density functional theory (DFT), it will be highly valuable to push the accuracy of MLFFs beyond the DFT level in the future.

Among the data distillation approach,
the recently proposed Random Network Distillation (RND) algorithm \cite{Finkbeiner2023GeneratingMT}
and the Dimensionality Reduction Encoding Clustering Tiered sampling (DIRECT) algorithm \cite{Qi2023RobustTO} stand out for their efficiency in pruning training data for downstream applications.
%have demonstrated significant efficacy. 
The RND algorithm employs a fixed target network and a trainable predictor network to select representative atomic configurations based on descriptor-embedded inputs. By identifying structures with the highest distillation error, the method ensures high structural diversity within the training set. 

The DIRECT algorithm facilitates the selection of robust training sets by encoding atomic configurations into high-dimensional descriptors and applying dimensionality reduction via Principal Component Analysis (PCA). 
The resulting space is partitioned into structural clusters using K-means, followed by stratified sampling to ensure comprehensive coverage of the configuration space, including rare outlier regions. 
Although these methods excel in single- and multi-phase systems, they are yet to be tested on phase transition regimes.

Aside from these two algorithms, and driven by the interest in studying phase transition problems, in this work we conduct a specific examination of configurational data distillation algorithms for MLFF model training in the liquid-liquid phase transition (LLPT) regime of dense hydrogen \cite{Wigner1935OnTP,mcmahon2012properties,pierleoni2016liquid,morales2010evidence,cheng2020evidence,Fang02012019,istas2025liquid}.
In this work, we introduce a Central-Peripheral Distillation (CPD) algorithm to train MLFF model in the LLPT regime of dense hydrogen. This method optimizes data distillation within complex configuration spaces that include phase transition regions, ensuring the model captures critical structural features in both stable phases and transition zones. 
Our results demonstrate that CPD not only substantially reduces the required training data but also ensures the stability and accuracy of the potential under extreme physical conditions, offering a robust tool for the study of complex phase behavior.
In Sec.~II, we describe the details of the CPD algorithm and other computational settings.
In Sec.~III, we present the performance of the algorithm on reducing the training dataset and the accuracy of the MLFF obtained.
Extended discussion and conclusions are included in Sec.~IV and V, respectively.

\section{Methods}

\begin{figure*}[t]
    \centering
    \includegraphics[width=1\linewidth]{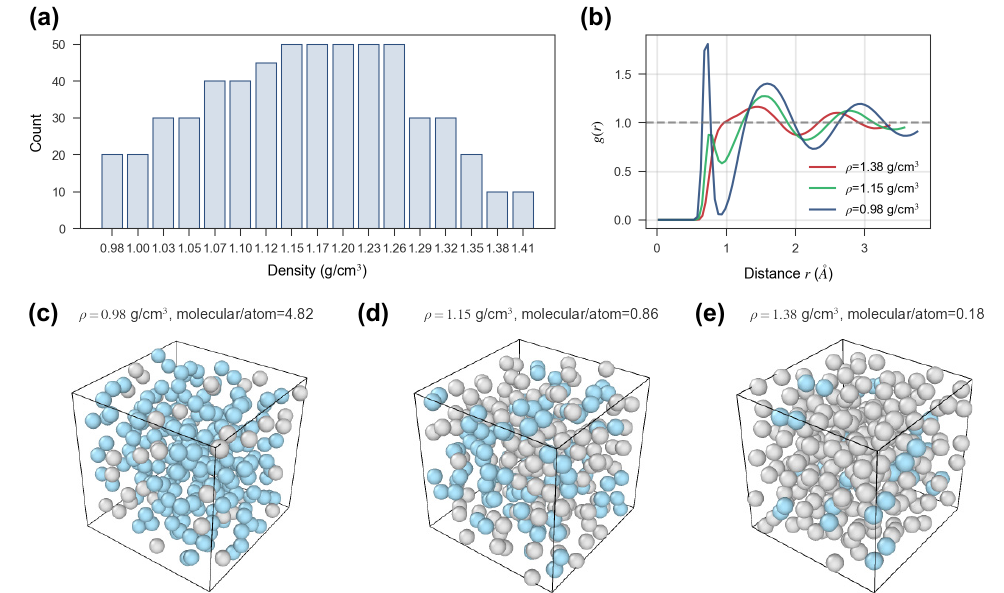}
    \caption{Structural characterization and phase transition of the hydrogen test dataset for LLPT at 1000 K. (a) Histogram of the density distribution for the 575 configurations, covering a range from 0.98 to 1.41 g/cm\textsuperscript{3}. (b) Radial distribution functions (RDFs) for the molecular($\rho=0.98g/cm^3$), transition($\rho=1.15g/cm^3$), and atomic regimes($\rho=1.38g/cm^3$), illustrating the structural evolution during the phase transition. (c--e) Representative snapshots at different densities, where hydrogen atoms in the atomic and molecular phases are colored light gray and light blue, respectively. (c) molecular phase, (d) transition region and (e) atomic phase.}
    \label{fig2}
\end{figure*}

\subsection{CPD workflow}
The CPD procedure begins by extracting features from molecular configurations with the MACE descriptor \cite{batatia2022mace}, which maps the atomic environments into a high-dimensional latent space. 
To reduce the dimension, PCA is applied to the resulting set of latent vectors, projecting them into a feature space.
In this reduced space, the local density $\rho_i$ for each data point $i$ is defined as the number of neighboring points located within a fixed cutoff radius $r_0$:
\begin{equation}
    \rho_i = \sum_{j \neq i} \mathbb{I}_{N(x_i)}(x_j)
\end{equation}
where $N(x_i) = \{ x \mid \|x_i - x\| \le r_0 \}$ is the neighborhood set of $x_i$ within a cutoff radius $r_0$, and $\mathbb{I}_A(\cdot)$ denotes the indicator function of a set $A$. The optimal value of $r_0$ is determined through an iterative process aimed at maximizing the variance of the density distribution while minimizing the count of isolated points (i.e., points where $\rho_i = 0$).

Based on the calculated density distribution, we implement a weighted sampling strategy. Specifically, sampling is strictly confined to the top $\alpha$\% densest and bottom $\beta$\% sparsest points. This $\alpha$ and $\beta$ \% threshold is selected to balance two constraints: it ensures the candidate pool exceeds the target size of the distilled dataset for valid sampling, while remaining sufficiently narrow to prevent degradation in the density-based classification performance. The rationale behind this dual-focus approach is rooted in the physical characteristics of phase transitions. Systems undergoing phase changes often exhibit significant structural fluctuations and produce numerous outliers near the transition boundaries. By assigning high weights to the densest regions, we effectively capture the representative features and typical structures of each phase. Moreover, selecting the sparsest regions allows the model to encompass critical outliers and rare configurations induced by the phase transition, thereby significantly enhancing the structural variety and robustness of the distilled dataset.

\subsection{HLLPT1k dataset}
For the purpose of this study, we constructed a new energy/force dataset for the liquid-liquid phase transition of dense hydrogen at 1000 K, dubbed as the ``HLLPT1k'' dataset.
To generate the structures of the dataset, we performed \textit{ab initio} molecular dynamics (AIMD) simulations using the Quantum ESPRESSO package \cite{giannozzi2009quantum,giannozzi2017advanced}. 
The initial atomic configurations, consisting of 256 hydrogen atoms in a cubic cell, were randomly generated. 
The system density was systematically varied by sampling the Wigner-Seitz radius $r_s$ from 1.240 to 1.400 Bohr in uniform increments of 0.010 Bohr. 
DFT calculations were conducted using the vdW-DF exchange-correlation functional to account for dispersion interactions \cite{dion2004van}, combined with Hamann-Schlüter-Chiang-Vanderbilt (HSCV) norm-conserving pseudo-potentials \cite{vanderbilt1985optimally}.
An energy cutoff of 80 Ry is applied for the planewave expansion. 
The AIMD simulations were carried out in the NVT ensemble at a temperature of 1000 K, controlled by a stochastic velocity rescaling (SVR) thermostat \cite{Bussi2007CanonicalST}. 
Each simulation trajectory was propagated for 10,500 steps with a time step of 0.48 fs, resulting in a total simulation time of roughly 5 ps per trajectory. 
At the AIMD stage, the convergence criterion of self consistent field (SCF) calculation was set at $10^{-6}$ Ry.
For the construction of the dataset, snapshots were extracted from the production phase of the trajectories by discarding the first 1.0 ps of each simulation to ensure that the system had reached thermal equilibration. 
Subsequent total energy and force calculations were performed using the same DFT setting with a raised SCF convergence criterion of $10^{-9}$ Ry.

After a series of training and testing to ensure an optimal balance between training efficiency and accuracy for the MLFF model, we finally constructed the HLLPT1k dataset consisting of a total of 575 configurations with each system containing 256 atoms. 
The structures within the HLLPT1k dataset covers a density range spanning from 0.98 to 1.41 g/cm\textsuperscript{3} along the 1000 K isothermal line.
For each density value, 10 to 50 configurations were randomly sampled as shown in Fig.~\ref{fig2}(a).
Three typical RDFs are shown in Fig.~\ref{fig2}(b) and their corresponding structures are visualized in Fig.~\ref{fig2}(c-e), where a molecular phase, an atomic phase and a mixed phase are shown for comparison.
Specifically, the low-density regime corresponds to the molecular phase, marked by a prominent intramolecular peak at $\sim$0.8~\AA~in the RDF. 
This feature is suppressed in the transition regime and eventually vanishes in the high-density atomic phase.
It highlights that the dataset effectively covers the phase transition regime between the molecular liquid and the atomic liquid.

\subsection{MLFF model training}
% The MACE model is used as the MLFF. 
We utilized training sets distilled via different methods (DIRECT, RND, CPD, Random) to fine-tune the MACE model, thereby evaluating the effectiveness of the distillation process. Specifically, we employed the MACE foundational model (the``medium-density Agnesi'' variant)\cite{batatia2022mace} as the starting point for transfer learning. To ensure loss convergence, the model was trained for 500 epochs with a mini-batch size of 2 (details in the Supplemental Material, Fig.~S1). The loss function was carefully balanced by assigning weights of 10,000 to the energy and 550 to the forces, respectively. We adopted a base learning rate of 0.01 to facilitate efficient convergence.

\section{Results}

\begin{figure*}[t] % h:当前位置, t:顶部, b:底部, p:独立一页
    \centering
    % width=\textwidth 确保图片和你的文字行一样宽
    \includegraphics[width=\textwidth]{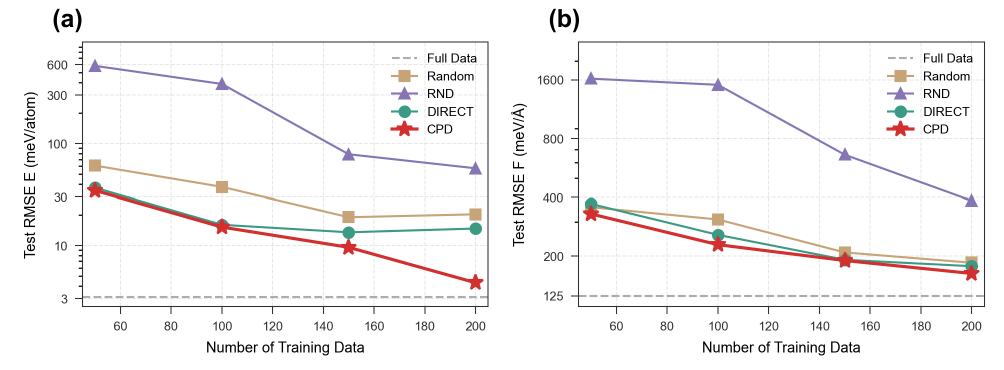} 
    
    \caption{Comparison of energy and force prediction performance. The RMSE of energy (a) and force (b) are plotted as a function of the number of training data selected using different data distillation methods. The dashed line indicates the error achieved with the full training dataset before distillation.}
    \label{fig3}
\end{figure*}

As demonstrated in Fig.~\ref{fig3}, we first evaluate the training performance of the proposed CPD algorithm in comparison with existing data distillation methods.
Here, the number of training data refers to the number of structures selected by the distillation algorithms to train the MLFF model. 
The root mean square error (RMSE) of energy and force are calculated using a test dataset of 95 structures.
%---including Random Network Distillation (RND) and Dimensionality Reduction Encoding Clustering Tiered-sampling (DIRECT)---as well as random sampling.
With CPD, the energy error decreases continuously from approximately 34.6 meV/atom with 50 training structures to 4.3 meV/atom with 200 training structures, already approaching the error level achieved with the full training dataset ($N=575$) which is 3.1 meV/atom.
The force RMSE also exhibits a steady decreasing trend, converging toward the value obtained with the full dataset.
Overall, in this test case, the CPD algorithm achieves predictive accuracy comparable to the full-dataset model while using only approximately 35\% of the total data, underscoring its efficiency for data distillation.

%the relative performance of the distillation algorithms shifts as the training set size increases. 
Among other distillation methods, only DIRECT delivers performance comparable to CPD in terms of the energy error when the number of structures is fewer than 100.
%configurations, the DIRECT and CPD algorithms yield nearly identical test losses, both significantly outperforming RND and random sampling. 
Beyond this point, the error of DIRECT plateaus and shows no further improvement with increasing training data size between 100 and 200.
The model converges to an energy RMSE of 14.7 meV/atom, which is 241\% larger than that obtained with CPD.
For comparison, we also performed random uniform sampling from the full dataset to fine-tune MACE with varying training set sizes. For this ``Random'' baseline, the energy error is consistently several times larger than the CPD result and plateaus beyond 150 training structures, further widening the performance gap relative to CPD.
In terms of force RMSE, “Random” and DIRECT exhibit comparable performance, only slightly inferior to CPD.
Somewhat surprisingly, RND displays the poorest performance, with substantially larger energy and force errors.
See Sec.~IV for more discussions on insights into the worse performances of these other algorithms. 

%However, a divergence emerges as the training set expands to 150 configurations. 
%While the test losses for random sampling and the CPD algorithm continue to decline, the performance of DIRECT plateaus, showing no substantial improvement over its 100-configuration baseline. 
%At this threshold, the CPD algorithm demonstrates a clear advantage, maintaining a lower test loss than all competing methods.
%Upon reaching a dataset size of 200 configurations, the test losses for random sampling and DIRECT stabilize, showing no further significant reduction. 
%Although the RND algorithm exhibits a decrease in loss at this stage, it remains substantially higher than that of its counterparts. 
%In contrast, the test loss of the CPD algorithm continues to diminish, reaching its minimum value and nearly converging with the performance of the model trained on the full dataset ($N=575$). 

%For force prediction (Fig.~\ref{fig:force_comparison}), the overall error is larger compared to energy due to parameter weighting settings. Nevertheless, the data selected by CPD consistently outperformed all other methods across datasets containing 50, 100, 150, and 200 configurations.

%Across all training set sizes of 50, 100, 150 and 200 configurations, the CPD algorithm consistently yielded lower test losses for both energy and force predictions compared to all alternative schemes under identical parameter settings.

It is also worth noting that our CPD algorithm has implemented a feature descriptor of the same type used in MACE, which may lead to a question of whether this is the main reason for the better performance.
To verify this, we conducted an additional investigation by replacing the MACE descriptor with a pre-trained SchNet descriptor \cite{Schtt2017SchNetA} (details in the Supplemental Material, Fig. S2).
We find the results remained consistent, that the CPD algorithm continued to outperform RND, DIRECT, and ``Random'', underscoring its robustness and generalization across different machine learning frameworks.

\begin{figure*}[t]
    \centering
    \includegraphics[width=1\linewidth]{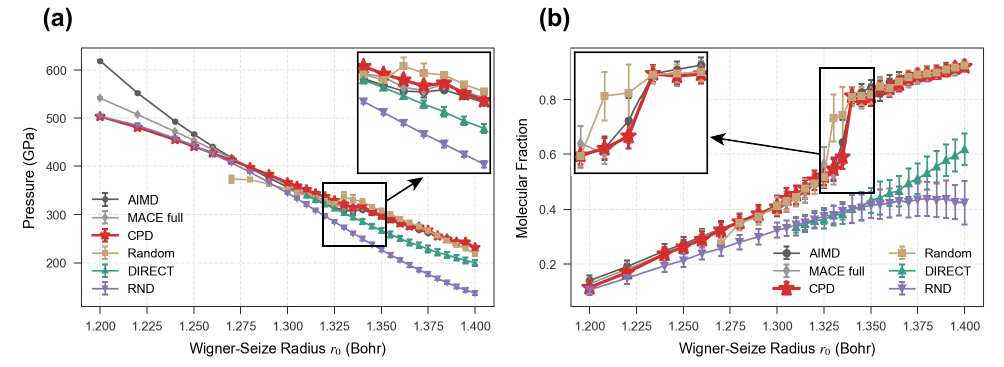}
    \caption{Performance of MLFFs trained on different datasets for hydrogen LLPT at 1000~K. Pressure (a) and molecular fraction (b) as a function of the density, labeled by Wigner-Seize Radius. The molecular fraction is defined following ref. \onlinecite{cheng2020evidence}, and a detailed calculation description in provided in SI.S3. The inset in (a) is a zoom-in of the phase transition regime. Models trained on DIRECT-distilled and randomly sampled datasets failed to yield reliable results for the first ten and five data points of the atomic phase, respectively. Consequently, these points are omitted from the figures.}
    \label{fig4}
\end{figure*}

To further evaluate the efficacy of the CPD algorithm, we assess the MD performance of the MLFF models beyond static energy and force metrics. Specifically, we compare models fine-tuned on 200 structures selected via different methods, alongside a reference model trained on the full dataset.
These models were subsequently employed to predict the liquid-liquid phase transition of hydrogen at 1000~K.
In Fig.~\ref{fig4} we benchmark 
the pressure and molecular fraction as a function of the density, labeled by the Wigner-Seize Radius, against AIMD results.

Our findings reveal substantial differences in how these models capture the complex thermodynamics associated with the hydrogen LLPT. 
As shown in Fig.~\ref{fig4}, the CPD based model accurately predicts both the pressure and molecular fraction, correctly reproducing the phase transition point as well as the slope of the curves in the transition region.
Notably, the CPD based model achieves accuracy comparable to the model trained on the full dataset while using 35\% training configurations.
These observations are consistent with the superior performance of CPD in energy and force prediction tasks.
Collectively, these results demonstrate that the CPD algorithm effectively selects the most physically informative configurations, enabling accurate modeling of complex phase behavior with a minimal computational cost.

For the other distillation methods, 200 structures are insufficient to yield a competitive MLFF model.
Among them,
the Random based model gives reasonable predictions for pressure and molecular fraction values in Wigner-Seize radius range between $r_s=1.270$ and $r_s=1.400$, but it clearly underestimates the phase transition point.
Furthermore, this model breaks down when performing MD simulations in the low $r_s$ regime.
For models trained on datasets from RND and DIRECT, the predicted pressure and molecular fraction are entirely inaccurate, and both methods fail to provide a physically meaningful description of the phase transition.
In contrast, the models trained on the full dataset and the CPD-distilled remain stable across all tested regimes, exhibiting a far superior ability to handle extreme and edge cases.

\section{Discussion}
Based on the presented results in Fig.~\ref{fig3} and Fig.~\ref{fig4}, the superiority of CPD over other distillation methods for phase transition becomes clear.
Unlike single-phase systems, systems undergoing phase transitions impose rigorous requirements on the training distribution of MLFFs. 

However, conventional distillation methods are often prone to failure when applied to datasets encompassing phase transition regimes. The performance of RND is heavily dependent on the initialization of its two randomly generated neural networks, leading to a lack of robustness in its results. Additionally, a strategy centered on high-variance data points renders the RND framework significantly susceptible to the influence of statistical outliers. In systems involving phase transitions, this tendency may result in the loss of essential structural characteristics and physical information. While RND was originally validated on single-phase systems, such as KCl and NaCl molten salts \cite{Finkbeiner2023GeneratingMT}, it exhibits suboptimal performance in the phase-transition systems examined in this study, failing to accurately identify phase transition points in terms of pressure and molecular proportions.

On the other hand, DIRECT demonstrates acceptable convergence on the loss curve, it exhibits significant deviations and instability in predicting pressure and molecular ratios. This discrepancy likely arises from the fact that DIRECT is optimized for large-scale datasets (e.g., the Materials Project, which comprises 1.3 million structures) \cite{Qi2023RobustTO}, whereas not suitable for small dataset for specific phase transition problem. Consequently, the distilled dataset generated by DIRECT remains excessively large, rendering it unsuitable for this problem.

In contrast, the CPD algorithm optimizes the sampling balance by leveraging the inherent interpolation strengths of machine learning models. Unlike RND, which is limited by its stochastic initialization and excessive focus on atypical outliers, CPD effectively incorporates both the central points, which establish a robust baseline for interpolation in stable regimes, and the peripheral points, which preserve the drastic structural shifts near the transition point. By comprehensively covering both the cores and the boundaries of the configuration space, the CPD approach ensures that the MLFF maintains high accuracy and robustness across the entire thermodynamic range.

\section{Conclusion}

In this study, we introduced the CPD algorithm, a data distillation scheme specifically designed for phase transitions. 
Our validation on the hydrogen LLPT system demonstrates that the CPD algorithm outperforms existing distillation methods such as RND and DIRECT, as well as random sampling. 
Key findings indicate that a model trained on a CPD-distilled dataset, comprising only 200 configurations can achieve predictive accuracy for energy and force nearly identical to that of a model trained on the full dataset (deviated by 1 meV/atom). 
Furthermore, CPD-trained models exhibit superior numerical stability and accurately predict the thermodynamic quantities compared to AIMD benchmarks.
This method provides a highly efficient and robust tool for improving machine learning force fields, especially in scenarios where labeling costs are high, e.g. high-level quantum chemical calculations beyond the standard DFT-level methods are required. 
Beyond the hydrogen LLPT, the CPD algorithm holds great potential for accelerating the discovery and characterization of materials in extreme conditions and complex phase-change processes.

\begin{acknowledgments}
This work was supported by the National Science Foundation of China under Grant No. 52541026 and No. 12334003. We are grateful for computational resources provided by the High Performance Computing Platform of Peking University. 
We thank Bingqing Cheng for sharing the source code for molecular fraction calculation.
\end{acknowledgments}

\clearpage % 换新页
% 添加补充材料的标题
\begin{center}
  \textbf{\LARGE Supplemental Material for ``Beyond redundancy: dataset distillation for machine learning force field in phase transition regime''} \\[0.5cm]
\end{center}

% 重置所有计数器
\setcounter{equation}{0}
\setcounter{figure}{0}
\setcounter{table}{0}
\setcounter{page}{1}
\setcounter{section}{0}

% 重新定义编号格式，加上前缀 "S"
\renewcommand{\theequation}{S\arabic{equation}}
\renewcommand{\thefigure}{S\arabic{figure}}
\renewcommand{\thetable}{S\arabic{table}}
\renewcommand{\thesection}{S\arabic{section}}
\renewcommand{\thepage}{S\arabic{page}} % 可选：如果希望页码变成 S1, S2

% --- 标题与作者信息 ---
\title{Supplemental Material for ``Beyond redundancy: dataset distillation for machine learning force field in phase transition regime''}

\maketitle
The source code and datasets required to reproduce the results of this study are openly available on GitHub at: \url{https://github.com/cryQAQ/Central-Peripheral-Distillation}.

\section{Training Loss Curves under Different Scenarios}
\label{sec:loss_curves}

In this section, we present the training and validation loss curves to verify the convergence of the models under different data selection strategies. Figure S1 illustrates the loss profiles over 500 training epochs for the full dataset (575 configurations) and various subsets (200 configurations each) generated via the CPD, DIRECT, and RND algorithms, alongside a purely random sampling baseline. As clearly observed across all subpanels (a–e), the models successfully reach stable convergence, ensuring the reliability of the trained force fields.

\begin{figure}[h]
    \centering
    \includegraphics[width=\textwidth]{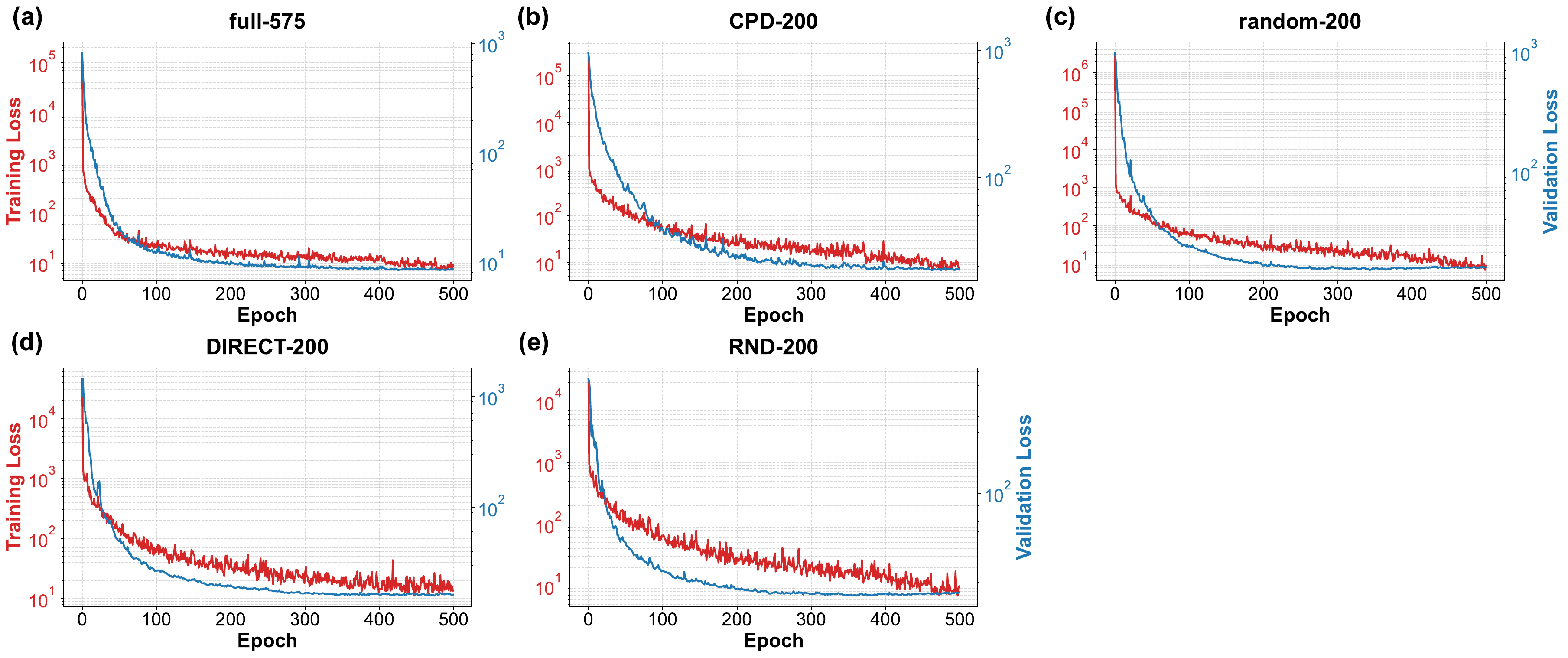}
    \caption{Evolution of training and validation losses over 500 epochs for machine learning force fields trained on various datasets. The subpanels illustrate the convergence behavior for: (a) the full reference dataset comprising 575 configurations; (b) the CPD-distilled dataset (200 configurations); (c) a baseline dataset comprising 200 randomly sampled configurations; (d) the DIRECT-distilled dataset (200 configurations); and (e) the RND-distilled dataset (200 configurations). All models demonstrate stable convergence within the allocated training epochs.}
    \label{fig:loss1}
\end{figure}

\clearpage % 换新页
\section{Comparison of Training Loss Using SchNet Descriptors}
\label{sec:schnet_comparison}

\begin{figure}[h]
    \centering
    \includegraphics[width=0.8\textwidth]{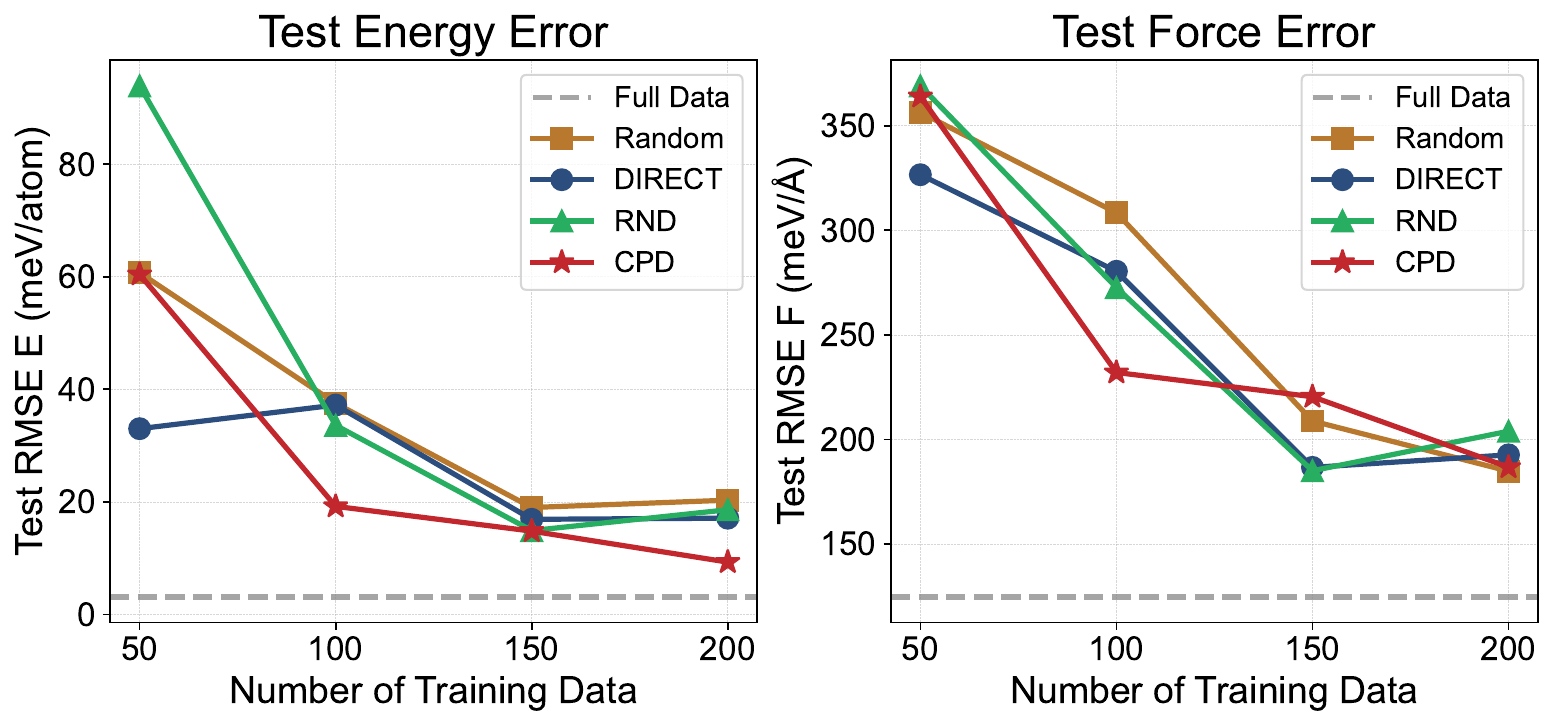}
    \caption{Comparison of final training loss utilizing the SchNet descriptor. The performance of different dataset distillation methods is plotted against varying sizes of the training set.}
    \label{fig:schnet_comparison}
\end{figure}

This section evaluates the performance of different dataset distillation methods. Figure \ref{fig:schnet_comparison} illustrates the final training loss utilizing the SchNet descriptor across different sizes of the distilled training sets. The comparison highlights the efficiency and accuracy of our proposed beyond-redundancy distillation approach compared to baseline methods.

\clearpage % 换新页
\section{Molecular Fraction Calculation}
\label{sec:molecular_fraction}
Following the approach proposed by Bingqing Cheng et al. \cite{Cheng2019EvidenceFS}, we evaluated the local coordination number (CN) for each atom along the molecular dynamics trajectories to quantify the molecular fraction of hydrogen. To mitigate artificial fluctuations caused by thermal vibrations at rigid cutoffs, a continuous cubic switching function was employed to define interatomic bonds. Specifically, the coordination contribution between an atom pair is $1.0$ for interatomic distances $r \le 0.8$~\AA, and smoothly decays to $0.0$ at a maximum cutoff of $r = 1.1$~\AA. The total CN of a given atom is the sum of these continuous fractional weights from all its neighbors. 

An atom is classified as part of a stable diatomic molecule (H$_2$) if its total CN falls within a Gaussian-smeared interval of $[0.8, 1.2]$. This continuous formulation intrinsically accounts for natural bond-stretching during thermal vibrations, while elegantly excluding both fully dissociated atomic hydrogen ($\mathrm{CN} \approx 0$) and transient polyatomic clusters or dense metallic states ($\mathrm{CN} \ge 2$). The final molecular ratio is derived from the statistical fraction of atoms satisfying this diatomic criterion.

To ensure reproducibility, the precise molecular fraction analysis was performed using the PLUMED code with the following configuration script.
\begin{lstlisting}
#UNITS LENGTH=A TIME=fs
COORDINATIONNUMBER ...
 LABEL=cn
 SPECIES=1-256
 SWITCH={CUBIC D_0=0.8 D_MAX=1.1}
 BETWEEN1={GAUSSIAN UPPER=1.2 LOWER=0.8 SMEAR=0.2}
... COORDINATIONNUMBER
 PRINT ARG=* FILE=COLVAR STRIDE=1 

\end{lstlisting}
\bibliography{ref}
\end{document}